# A New Approach to Image Compression in Industrial Internet of Things

Nahid Hajizadeh [1*], Pirooz Shamsinejad [2], Reza Javidan [3]

[1,2,3] Department of Computer Engineering and Information Technology, Shiraz University of Technology, Shiraz, Iran

* hajizadeh.nahid@gmail.com

**Abstract:** Applying image sensors in automation of Industrial Internet of Things (IIoT) technology is on the rise, day by day. In such companies, a large number of high volume images are transmitted at any moment; therefore, a significant challenge is reducing the amount of transmitted information and consequently bandwidth without reducing the quality of images. Image compression in sensors, in this regard, will save bandwidth and speed up data transmitting. There are several pieces of research in image compression for sensor networks, but, according to the nature of image transfer in IIoT, there is no study in this particular field. In this paper, it is for the first time that a new reusable technique to improve productivity in image compression is introduced and applied. To do this, a new adaptive lossy compression technique to compact sensor-generated images in IIoT by using K-Means++ and Intelligent Embedded Coding (IEC) as our novel approach, is presented. The new method is based on the colour of pixels so that pixels with the same or nearly the same colours are clustered around a centroid and finally, the colour of the pixels will be encoded. The experiments are based on a reputable image dataset from a real smart greenhouse; i.e. KOMATSUNA. The evaluation results reveal that, with the same compression rate, our approach compresses images with higher quality in comparison with other methods such as K-means, fuzzy C-means and fuzzy C-means++ clustering.

## 1. Introduction

Industrial Internet of Things (IIoT) is one of the most important and widely used areas in the Internet of Things. Using IIoT in industrial units, all the objects can be connected at the same time and create an integrated network to carry out all information exchange, control and monitoring tasks. IIoT consists of many devices that are connected by communication softwares. These devices can collect, control, exchange and analyse information in order to intelligently modify their behaviour or environment, entirely without human intervention. IIoT has vast applications in large industries such as industrial automation and power stations. Industrial automation is one of the most notable and widespread application of IIoT. By automating machines, companies able to operate in such a way that, applying high-tech software platforms, monitoring and controlling to make evolution in the subsequent production process be more impressive. In manufacturing automation, the major part of sensor networks is image sensors. Image sensors can detect and send the information that forms an image and hence plays a crucial role in security and safety applications. The operation of image sensors embedded in industrial automation which is equipped with IIoT is depicted in Figure 1. Each image sensor, at any moment, sends multiple images to the central processing system. Often, industrial automation devices are equipped with advanced and high-quality cameras that capture images at the desired resolution. To reduce the transfer rate, images can be sent in low quality so that processing cannot be performed correctly on that image. A solution that reduces image size without sacrificing its quality is using a compression technique.

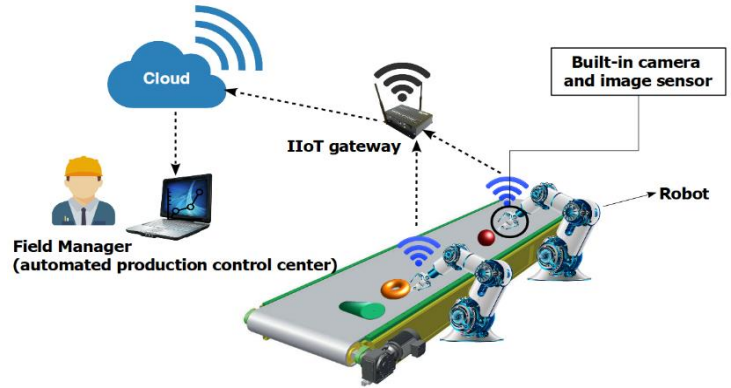

*Figure 1: IIoT in industrial automation*

This technique works on the basis of the coding system in such a way that pixels with the same or nearly the same colours are clustered around a centroid and finally the colour of the pixels will be encoded. In fact, the code has a much lower volume than the original image. To address the mentioned issue, in this research, an image compression technique to compress sensor-generated images in IIoT by using $K$-Means++ and IEC (Intelligent Embedded Coding) approach is developed. In this method, the number of colours presented in an image will be reduced to $K$ colors. These $K$ colors are centres of clusters in our clustering algorithm and consist of the pixel colours that occur most in the image.

The reason behind the use of fuzzy logic is the pixel colours property. For example, the orange colour is a combination of red and yellow colours, and if we consider red and yellow as clusters centres, the orange colour will have a degree of membership to the red cluster and a degree of membership to the yellow cluster.

The existing methods have not considered the particular feature in sensor-generated images in IIoT, and this feature is that the consecutive captured images have small changes compared to the previous submitted images, and this feature can be used to increase productivity. The purpose of this article is to explain this feature clearly and use it to promote efficiency.

In this research, a novel lossy compression technique to compress sensor-generated images in IIoT by applying K-Means++ and IEC approach is presented. The new approach is based on the colour of pixels, so those pixels with the same or nearly the same colours are clustered in a centroid and finally the colour of the pixels will be encoded. This study is based on an image dataset provided in a real smart greenhouse.

The main contributions of this paper are:

- The **first contribution** of this research is applying $K$-Means++ for image compression. As indicated in the related works section, this is for the first time that $K$-Means++ is used for compressing images. In addition, in this paper, the performance of fuzzy $C$-Means++ algorithm, in comparison with other methods, is delivered and discussed.
- The **second contribution** is related to the application of our approach in IIoT. So far, there has been no research on the compression of sensor images on IIoT. The dataset used in this research is quite relevant to industrial fields and provided in a real smart greenhouse.
- The **third contribution** is introducing an algorithm (e.g. IEC) to send images based on the volume of image changes. In this technique, the estimation time for transferring images depends on how much the new image has deviated from the base image. This is for the first time that this technique has been applied in IIoT.
- The **last contribution** but not the least one is that our method is designed and tested for both RGB and grayscale images. However, in the aforementioned literatures, only grayscale images have been analysed. Since some image cameras are RGB and need to be analysed in RGB code, we should use RGB images. Moreover, to compare the performance of K-Means++ algorithm to the other algorithms, grayscale images are also analysed.

The reminder of this article is structured as follows: In Section 2, some related works are reviewed. In Sections 3 the proposed approach is introduced. In Section 4, the experimental results and, consequently their evaluations are presented. We conclude our research in Section 5.

## 2. Related Works

Analysing the previous researches in image compression in wireless sensor networks can be categorised into two groups. In this part of the current section, a review of researches in image compression in wireless sensor networks, in general, is presented.

[1] is one of the first studies in the field of image compression in WSN. To make encoding and decoding faster, a method which its name is SPIHT (Set Partitioning In Hierarchical Trees) was introduced. SPIHT makes a tree structure based on the pixel data of the image and collects the most related pixels in a single set. After that, zero-valued trees are encoded in a single value. Applying this feature of wavelet transform and SPIHT leads to a high compression ratio in images.

Discrete Cosine Transform (DCT) is an image compression method that often is applied in WSN. DCT based image compression algorithms do compression in an effective way [2].

Sending image segments instead of the whole image is introduced in [3]. This method at first determines the object presence and then sends image segments. The receiver constructs the whole image based on the segments that he received. The researchers claimed that their approach is practical in real-time object tracking. However, they did not reveal any detail of their experiments such as the distance between the nodes, the environment area, number of hops, etc.

Compression of video data in WSN has been analyzed in [4]. In this approach, redundant features between the frames are eliminated to reduce the bit rate of the video stream.

In the following part of the current section, a review of researches in image compression particularly based on the K-Means clustering algorithm is presented.

Using K-Means-based clustering algorithm for image compression has been used by Paek et al. [5] to compress images in wireless sensor networks. They focused on the similarity of pixel colours to aggregate pixels and compress the input image. Their real-world data set includes 100,000 images that were collected from their pilot deployment. They found that their proposed approach could reduce power usage by 49%, approximately, if the learning phase is done offload and just do the compression step.

In [6], they extended their prior work and proposed an approach called H-K compression, a simple, lightweight grayscale image compression algorithm which is the integration of Huffman coding and K-means clustering. Their proposed algorithm works in a way that reduces the computation cost, by applying $k$-means clustering for aggregating the pixel colours and Huffman coding technique for encoding the group colours. Their evaluation revealed that their proposed algorithm compresses the images by 57% and reduces the power consumption by 43%.

The same authors later in [7] introduced an energy-efficient method using image resizing and colour quantization, which reduce the amount of data transferred. They showed that if an image is highly compressed, it can be well-classified with a Convolutional Neural Network (CNN) model. In fact, they used deep learning tools to cluster images. They noted that high-resolution raw images captured with low-power and limited battery cameras can be clustered by a CNNs model. Their evaluation illustrated that if they can maintain the accuracy of classification to 98%, transmitted data will be reduced by 71%, which means that energy consumption will be reduced by 71% too.

Heng et al. [8] proposed a distributed image compression architecture over a wireless multimedia sensor network to improve the overall network lifetime. Their simulation results revealed that their method could extend the overall network lifetime and achieve high throughput.

Sheeja and Sutha [9] introduced a new wireless sensor network-based disaster rescue telemedicine strategy to maximize network lifetime and minimize energy consumption. They compounded three central concepts to achieve their goals which is expanding the lifetime of a WSN; node clustering, medical image compression, and node energy management. In addition, they introduced three novel algorithms, which are named 'Network clustering using Non-border CH oriented Genetic algorithm, Fuzzy rules and Kernel FCM (NCNBGF)', 'High gain MDC algorithm (HGMDC)' and 'Critical node handling using job limiting and job shifting (CJLS)'. In their simulation results, their proposed method reaches to 43% energy saving while other methods achieve 20% energy saving in total.

Kaljahi et al. [10] introduced a novel image size reduction model for energy-efficient transmission in Visual Sensor Networks (VSNs) by finding overlapping regions using camera location, which gets rid of unfocussed regions from the input grayscale images.

Our method is completely novel in its field and is based on a new lossy compression technique to compress sensor-generated images in IIoT, which uses K-Means++ and IEC approach presented in this article. Moreover, unlike existing techniques that consistently compress and send images, this technique works optimally and only compresses and transmits an image when it differs from the previous upload image.

## 3. The Proposed Approach

The proposed method contains two main subsections; Intelligent Embedded Coding and Image Compression by K-Means++ and fuzzy C-means Clustering.

### 3.1. Intelligent Embedded Coding (IEC)

The term "intelligent" is stand for this purpose that the process of encoding depends on some circumstances. In this case, encoding relies on the volume of image changes. In [5] some experiments are done to estimate the time to coding and sending the images. times, images do not have significant differences. distortion threshold and to gain higher performance, they considered one-week interval. other methods (i.e. predefined fixed intervals) is that IEC is not limited to a specific case but it is general and practical for any case. Other methods, by contrast, are completely depend on the situation of the cases. The structure of IEC algorithm is Nevertheless, our encoding interval is not They tried to recognize a fixed appropriate periodic interval for making compressed images. Finally, according to the average specified but is flexible. The superiority of our technique shown in Figure 2. In our method, transfer rate is proportional to the amount of changes in images. The time an image, in comparison with the stored image in a sensor, does not have significant changes (significant change is determined through a threshold value), no images will be sent to the server. Our strategy has two major benefits. First, the transfer rate will be reduced, because it is not needed that every image being captured be sent to the server. Second, it helps the control unit to quickly discover that some changes to the status of under control items have been occurred, because an image is sent to the server and subsequently to the control unit when

```
Algorithm: IEC
Input ← new_image, stored_image, threshold_value

while stopping criterion has not been met
do new_image ← take a new image
   Extract_similarity between new_image and stored_image
   if similarity < threshold_value
      Encode (new_image)
      Send (new_image)
      stored_image ← new_image
   end if
end while
```

*Figure 2: Intelligent Embedded Coding algorithm*

significant changes were made to the saved image. The flowchart of IEC approach is presented in Figure 3.

### 3.2. Image Compression by K-Means++ and fuzzy C-means Clustering

Before going through our new method, a brief introduction to the algorithms investigated in this article is presented.

The term "$K$-means" was introduced for the first time by James MacQueen in 1967 [11]. It is the most significant flat clustering algorithm and its purpose is partitioning $n$ members into $k$ clusters in such a way that the average squared Euclidean distance of members from their cluster centers be minimized. In this case, cluster centres have the greatest distance to each other. The most appropriate number for clusters ($K$) is unknown at the beginning of the algorithm and should be specified according to the data. Fuzzy $C$-means (FCM) clustering was introduced by J.C. Dunn in 1973 [12] and J.C. Bezdek improved it in 1981 [13]. In non-fuzzy clustering, members are related to separate clusters, where each member can only belong to one cluster, however in fuzzy clustering, members can belong to multiple clusters. To each of members, a membership degree is assigned. These membership degrees represent the degree to which members belong to each cluster. K-Means++ Clustering was proposed by Arthur et.al. [14] as an algorithm for choosing the seeds for the $K$-means clustering algorithm.

As a matter of fact, this algorithm tries to improve the initialization step of the $K$-means clustering algorithm by finding cluster centres that minimize the intra-class variance (i.e. the sum of squared distances from each data point to its cluster centre).

### 3.3. Why K-Means++ algorithm?

As many scholars have noted in their researches [15] [16] [17] [18], K-Means clustering algorithm is one of the most successful clustering algorithms. Moreover, choosing appropriate initial values for centroids in K-Means clustering algorithm has always been a challenging issue and many researchers have been trying to improve this algorithm in terms of determining better initial centroids [19] [20] [21] [22] [23]. The reason for choosing K-Means++ algorithm in the current article is the remarkable superiority of this algorithm over the K-Means clustering algorithms. The creators of this algorithm have introduced a novel strategy to seed the K-Means algorithm.

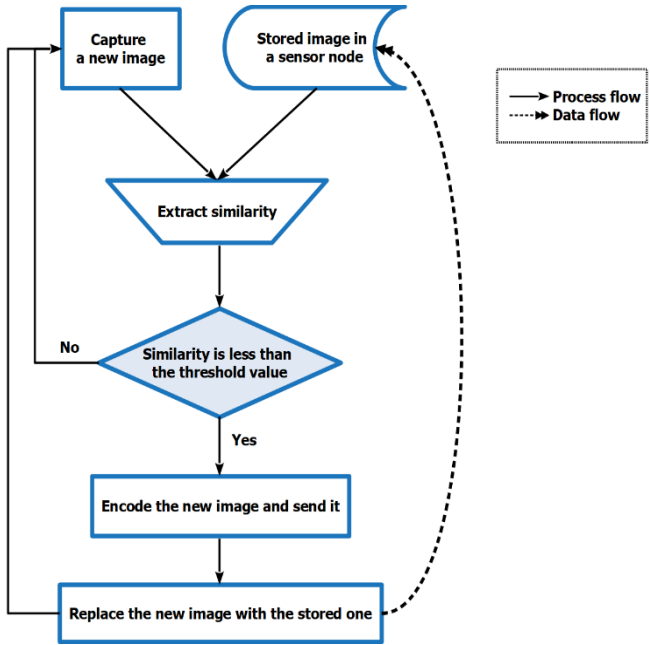

*Figure 3:* The flowchart of IEC approach

*Table 1*: the compression ratio for different values of K

| K | 4 | 8 | 16 | 32 |
|---|---|---|---|---|
| Bits per pixel | 2 | 3 | 4 | 5 |
| Compression Ratio | 4 | 2.67 | 2 | 1.6 |

The seeding method is fully applicable because it is very fast and simple. The results of experiments on real-world databases show that K-Means++ considerably outperformed standard K-Means in speed and accuracy.

## 4. Experimental Results

In this section, some experiments to evaluate the performance of our image compressor on a real-world dataset are represented. It is compared with 3 other algorithms; fuzzy $C$-means, $K$-means, and fuzzy $C$-means++ clustering algorithms. For this purpose, the investigated algorithms are implemented and executed on a real industrial dataset.

### 4.1. Metrics

There are prevalent metrics to compare two images and measuring the difference between lossy compressed and uncompressed images.
The assessments are carried out based on the famous metrics which are broadly used to compare the quality of images. These metrics along with their definitions and formulas are listed in Table 2.

### 4.2. Compression Ratio

In our image processing method, the output image only includes the colours that are more common in that, hence the number of colours to present an image is decreased. Our image compression technique is demonstrated in Figure 4.
In our datasets, images have different sizes, so our method is not limited to a specific size. Each image is represented by $n*n$ pixels and each pixel is indicated by an 8-bit unsigned integer which, depending on the image that is color or grayscale, determines the colour intensity.
As the colour spectrum is expanded from 0 to 255, so by decreasing the 256 colours of the images to only $K$ colors, the size of compressed images will be reduced too. The encoding technique is based on the similarity between pixel colours and the colours of $K$ cluster centers.
The number of elements in the middle image matrix are the same as input image but their elements values are the codes of $K$ centroids. In a real situation (e.g. an industrial environment) image sensors compress the taken image and then send the middle image and a vector of $K$ codes to a destination. After transmitting, the receiver decodes the images and reconstruct the original image by using middle image and the code word vector.
As mentioned above, any image has $n*n$ pixels and any pixel is a 8-bit unsigned integer. So, the size of an image is $n*n*8$ bits. By applying our fuzzy image compression, each pixel can be presented by only $\log_2 K$ bits. In this case, for an image consist of $8*n^2$ bits, there is needed to consider $\log_2 K$ bits plus a vector size in length $K$, that each element is 8 bits, to save pixel colors of centroids. The compression ratio [24] is shown in eq.(1).

$$Compression\ Ratio = \frac{Input\ (Uncompressed) Image\ Size}{Output\ (Compressed) Image\ Size}$$
$$= \frac{8n^2}{((\log_2 K)n^2 + 8K)} \quad (1)$$

For a large amount of $n$, the eq.(1) approximates to $\frac{8}{\log_2 K}$. To be more tangible, the compression ratio for a given number of $K$ is calculated and listed in Table 1.

### 4.3. Dataset

The image dataset used in this research is KOMATSUNA dataset [25]. One of the purpose of preparing this dataset is measuring plant growth and environmental information in indoor ecology. It contains RGB images of Komatsuna (i.e. a Japanese leaf vegetable). The suppliers of this dataset constructed a platform to capture plants from the top view. They cultivated five Komatsuna plants with 24 hours lighting (to accelerate the growth of plants) and captured every 4 hours for 10 days. The Komatsuna dataset platform and its RGB cameras are shown in Figure 5 (a) and (b), respectively. In Figure 6, a plant growth sample in a two-day period, in RGB and grayscale images, is depicted.

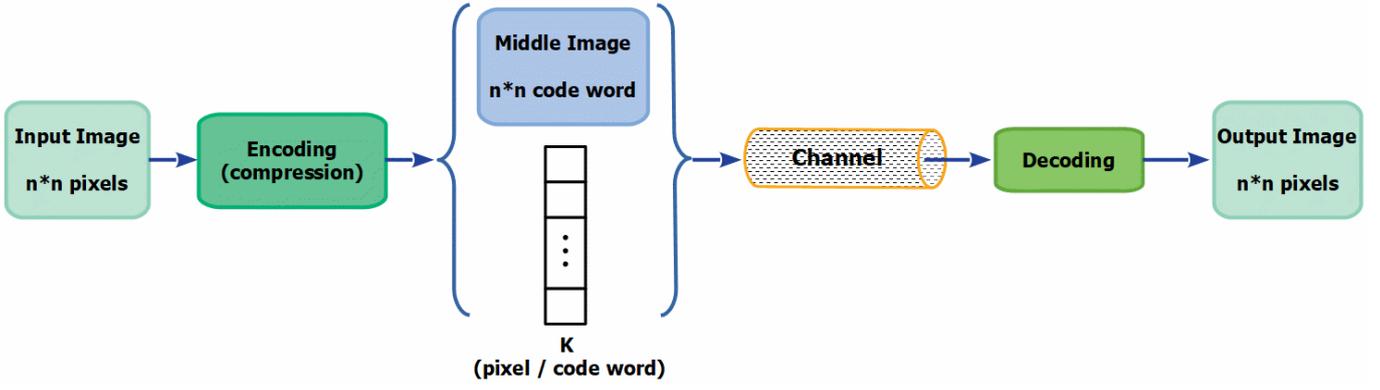

*Figure 4: An overview of our compression process*

*Table 2: The metrics and their definitions and formulas*

| Metric | Definition | Formula |
|---|---|---|
| MSE (Mean Square Error) | It measures the average of the squares of the errors between each original pixels and compressed pixels. | $MSE = \dfrac{1}{m*n}\sum_{i=0}^{m-1}\sum_{j=0}^{n-1}[I(i,j)-K(i,j)]^2$ |
| PSNR (Peak Signal-to-Noise Ratio) | It is the ratio between the maximum possible power of a signal and the power of corrupting noise. PSNR is defined via the MSE. | $PSNR = 10 * \log_{10}\left(\dfrac{MAX_I^2}{MSE}\right)$ |
| SSIM (Structural SIMilarity) | It is a method for measuring the similarity between two images. This measurement or prediction of image quality is based on an initial uncompressed or distortion-free image as reference. SSIM is designed to improve on traditional methods such as PSNR and MSE. | $SSIM(x,y) = \dfrac{(2\,\mu_x\mu_y + c_1)(2\,\sigma_{xy}+c_2)}{(\mu_x^2+\mu_y^2+c_1)(\sigma_x^2+\sigma_y^2+c_2)}$ |

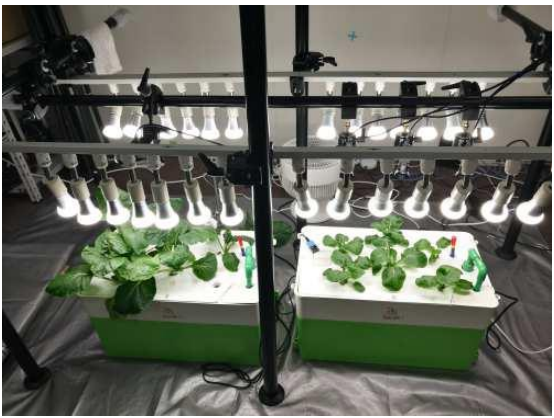

(a) The Komatsuna dataset platform

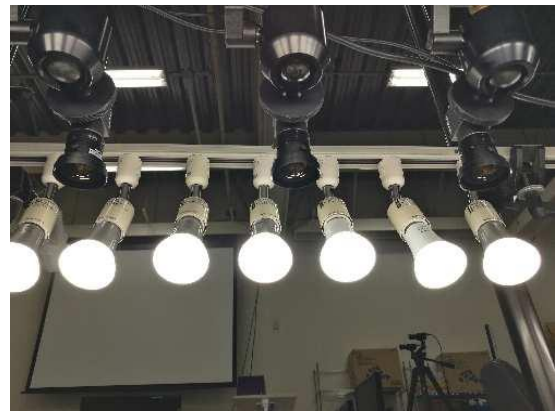

(b) The RGB cameras for capturing from a short distance.

*Figure 5: KOMATSUNA greenhouse*

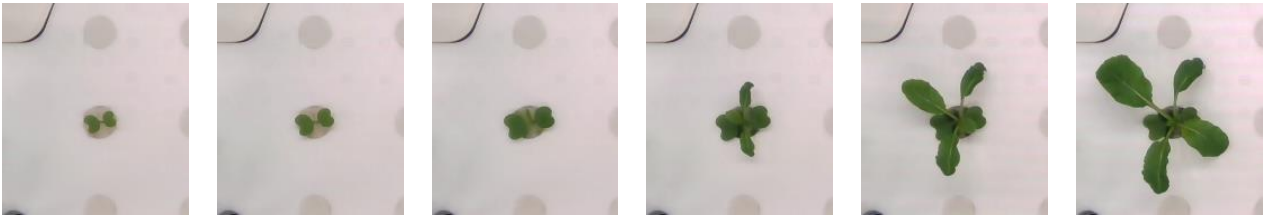

(a) A plant growth in RGB images

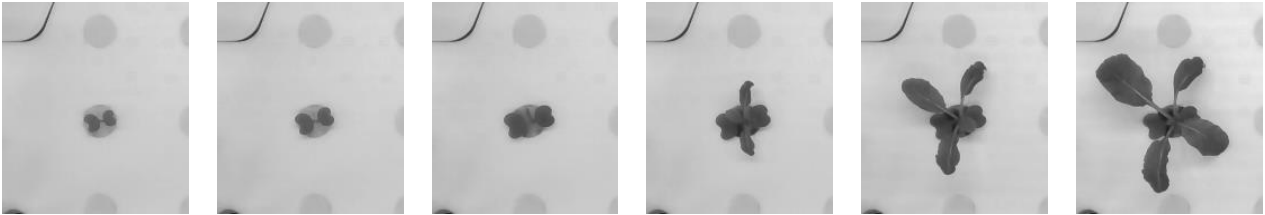

(b) A plant growth in grayscale images

*Figure 6:* *A plant growth in a two-day period*

This dataset is an extremely appropriate example of a smart greenhouse based on IIoT.

A smart greenhouse can control the environmental variables which affect the growth of crop. The most common of these controls include temperature, humidity, lighting and nutrient control which each of them has special sensors to sense them. In smart greenhouse based on IIoT, the information obtained by sensors is aggregated in an IIoT gateway and then send to a server. The unit controller by using this information can issue commands to meet the current needs of plants. This process for a special sensor (i.e. image sensor) is illustrated in Figure 7.

### 4.4. The impact of centroid updates

In this subsection the impact of centroid updates on the quality of output images is analysed.

In Figure 8 the MSE amount for each compressed plant image in two states is measured; the first state, when an arbitrary selected plant image (in this case img_5) is applied for learning and determining the centroids for compressing other plants images, and the second state, when each plant image is compressed based on the centroid determined per image.

According to Figure 8, when an image is capture from similar to img_5, they have some similar content, so the amount of distortion is decreased. However, images that have more difference with img_5, have higher degree of distortion. This issue implies that plant images can be entirely distinct, therefore, using a single plant image to extract centroids for compressing other plant images is not reasonable. For this reason, to achieve the best compression rate, every centroid set should be obtained per image.

### 4.5. The impact of *K* values

To achieve a high performance in *K*-means clustering and its extended algorithms, selecting an appropriate value for *K* is a crucial issue.

According to compression ratio section and eq.(1), for more compression ratio, the *K* value has to be lower, because, in this case, the bits needed to store an image is decreased. But this matter sacrifices the quality of images. On the other hand, a large value for *K* causes more color codes be available which leads to images with higher quality but lower compression ratio. Hence, for selecting a *K* value, there is a tradeoff between compression ratio and quality of images.

Due to the fact that the majority of previous researches applied an special range of *K* values (i.e. 4,8, and 16), consequently, in this article, the popular range plus additional *K* value (i.e. 32) is considered.

Another reason for selecting these values for *K* is that *K* values less than 4 causes a very low quality image and *K* values greater than 32 leads to compression ratio lower than 2 which is not effective.

In Table 3, six sample images, 3 RGB and 3 grayscale, compressed by K-Means ++ with four K values along with their RMSE, PSNR and SSIM values are presented. According to Table 3, clustering with k=32 and k=16 indicates very slight degradation in image quality.

The images compressed with k=32 have a very little visual differences in comparison with the input images.

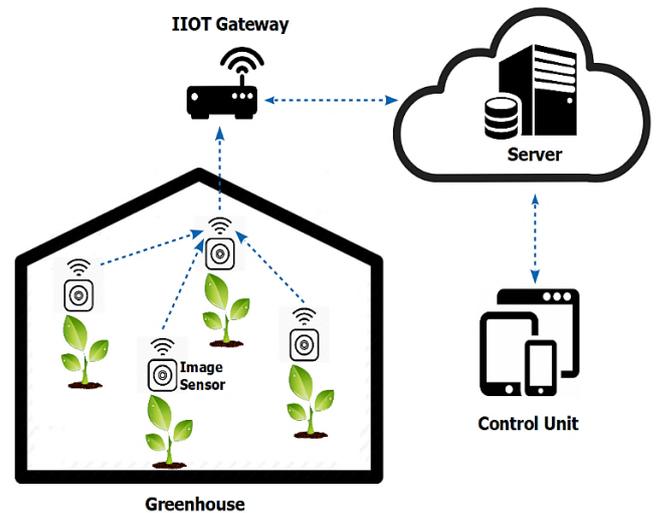

*Figure 7:* *A smart greenhouse controlled by IIoT*

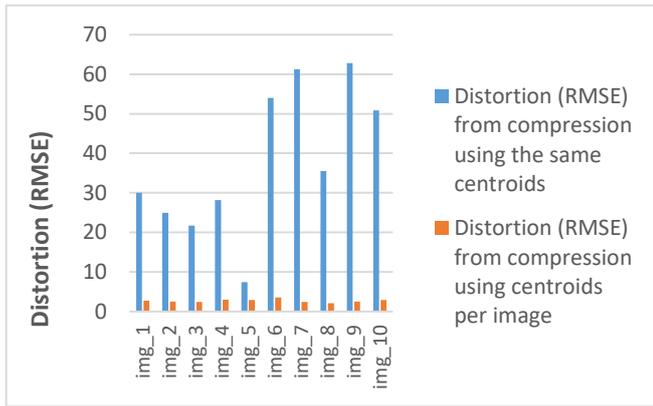

*Figure 8:* Distortion (RMSE) from compression using the same centroids or centroids per image

*Table 3:* sample images compressed by K-Means ++ with four K values

| Input image | K=32 | K=16 | K=8 | K=4 |
|---|---|---|---|---|
| | RMSE=1.3076 PSNR=45.8015 SSIM=0.9927 | RMSE=1.8253 PSNR=42.9039 SSIM=0.9890 | RMSE=3.0418 PSNR=38.4683 SSIM=0.9732 | RMSE=5.9419 PSNR=32.6523 SSIM=0.9641 |
| | RMSE=1.4477 PSNR=44.9196 SSIM=0.9912 | RMSE=2.7669 PSNR=39.2910 SSIM=0.9746 | RMSE=5.1068 PSNR=33.9678 SSIM=0.9516 | RMSE=7.1165 PSNR=31.0854 SSIM=0.9418 |
| | RMSE=1.8296 PSNR=42.8836 SSIM=0.9893 | RMSE=3.4642 PSNR=37.3388 SSIM=0.9702 | RMSE=6.2181 PSNR=32.2576 SSIM=0.9310 | RMSE=11.3032 PSNR=27.0668 SSIM=0.8944 |

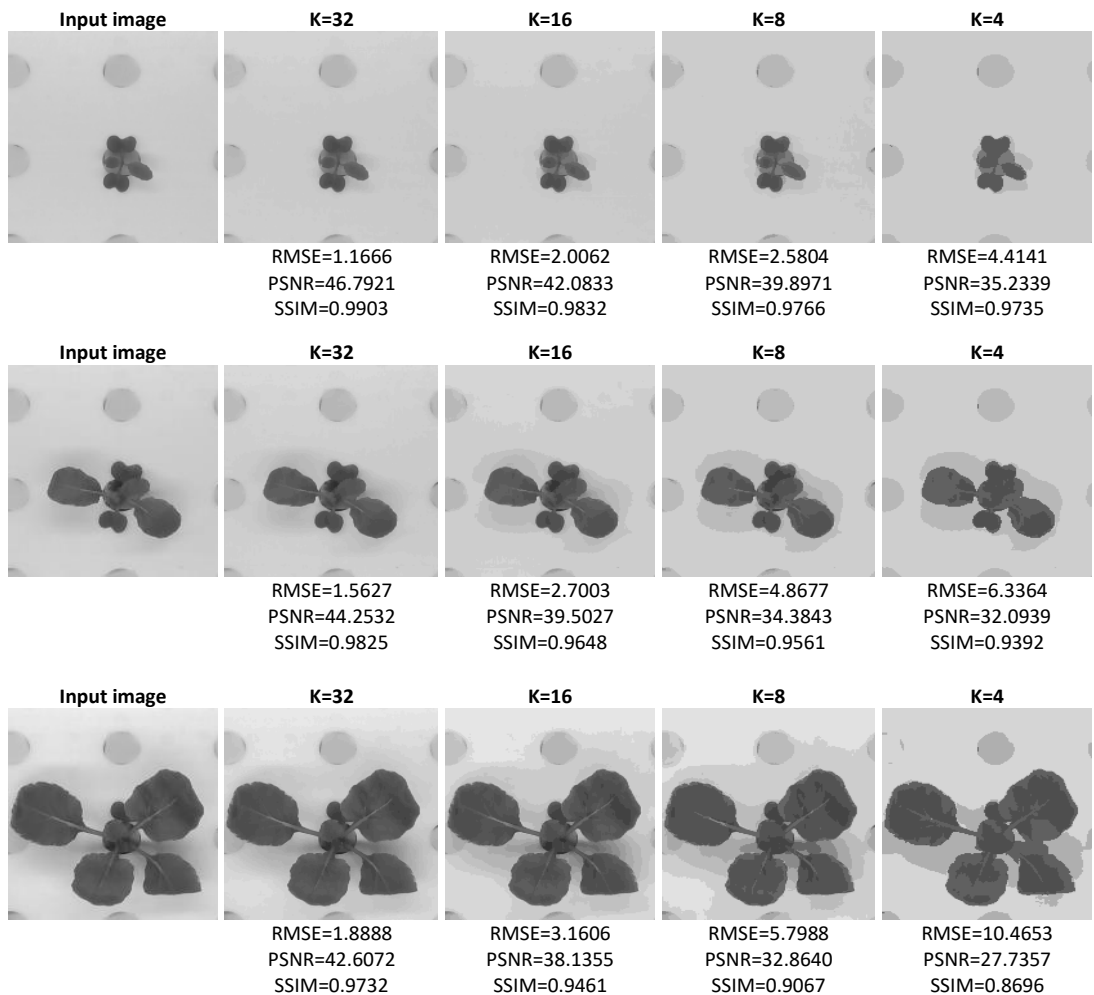

The amount of distortion (i.e. RMSE) for the mentioned compressed images based on different k values are plotted in Figure 9.

In this research, four algorithms (i.e. K-Means, K-Means ++, fuzzy C-means, and fuzzy C-means++) that each of them is implemented for both RGB and grayscale images are taken into consideration.

To find out whether there is a significant difference between the aforementioned algorithms, three box plots, based on the metrics, for any of the algorithms are prepared. To achieve the box plots, 30 independent runs for any algorithm are done. In Figure 10, the algorithms are compared according to their RMSE values.

The lower RMSE shows the better performance, so K-Means ++ surpasses the other algorithms at this stage of the competition. Fuzzy C-means algorithm has nevertheless won the last place.

From SSIM point of view, higher values are more appropriate. According to Figure 11, K-Means ++ achieved the best SSIM and ,on the contrary, Fuzzy C-means algorithm gained the worst SSIM.

Like SSIM, higher value for PSNR is more desirable.

According to Figure 12, K-Means ++ also defeats the other participants and fuzzy C-means algorithm is still in the last row.

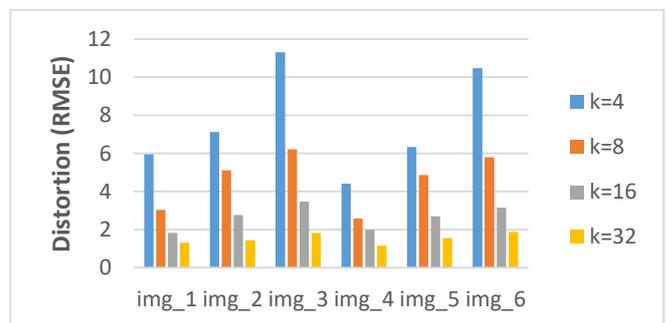

*Figure 9: The amount of distortion (RMSE) for different k values*

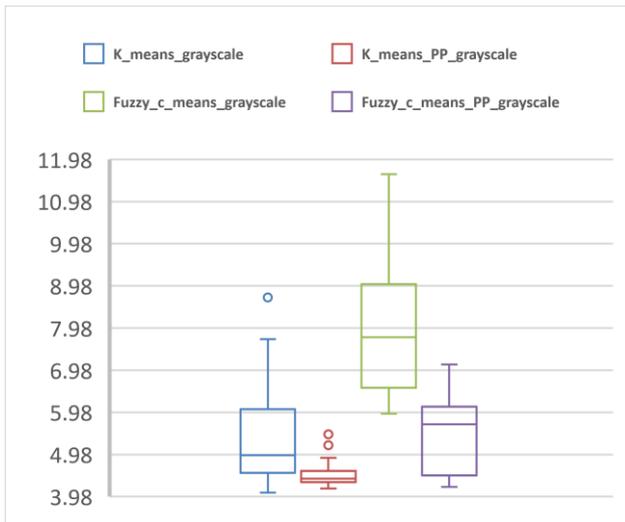
RMSE for grayscale images

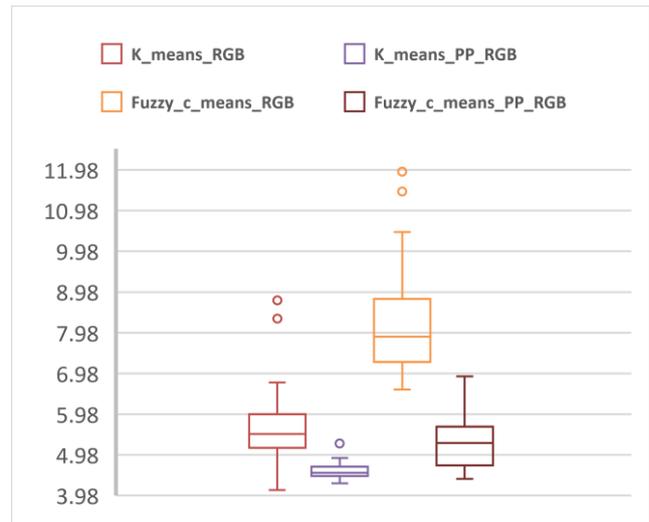
RMSE for RGB images

*Figure 10:* *The box plot of RMSE values for the aforementioned algorithms*

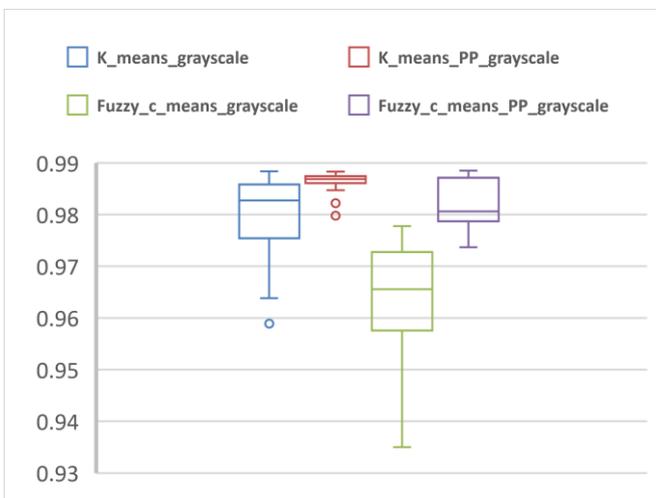
SSIM for grayscale images

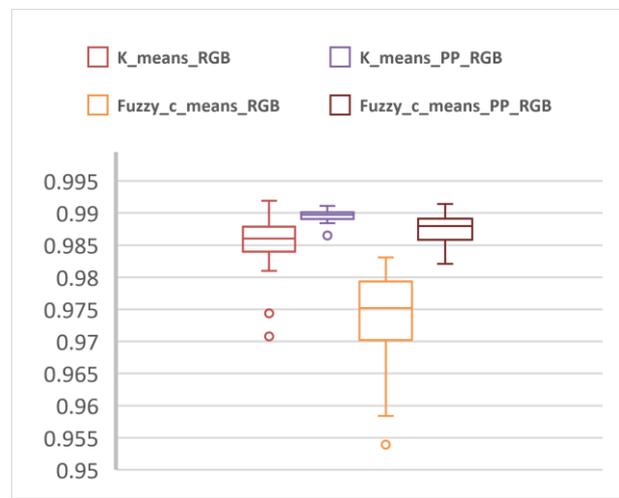
SSIM for RGB images

*Figure 11:* *The box plot of SSIM values for the aforementioned algorithms*

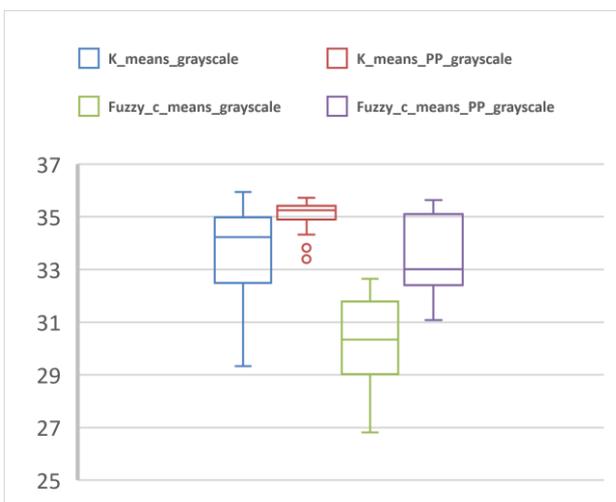
PSNR for grayscale images

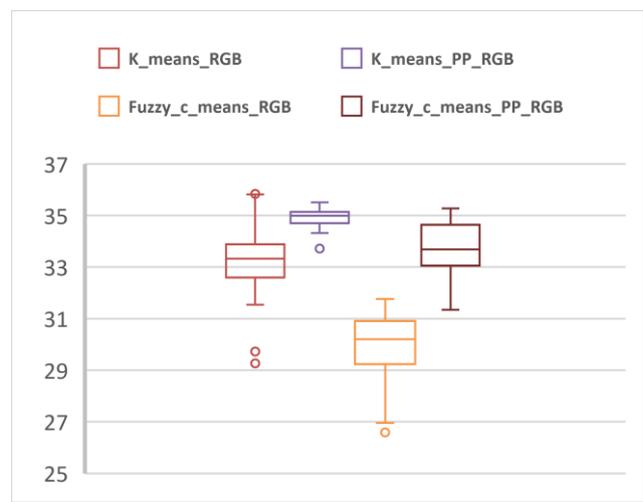
PSNR for RGB images

*Figure 12:* *The box plot of PSNR values for the aforementioned algorithms*

## 4.6. The impact of tonal distribution

To analyse the effect of tonal distribution on the performance of fuzzy algorithms, an experiment is done on three images with different tonal distribution and their histograms (a graphical representation of tonal distribution in a digital image [26]) are plotted.

In two images, the tonal distribution for each colour is the same, however in another image, there are different colours with different tonal distribution. These three images along with their image histograms are illustrated in Table 4.

As is clear in Table 4, the image histogram of green and colour leaves are the same, however, plants in the greenhouse. shows diferent image histogram The reason for this is that the colors in the image are not uniformly distributed. This hypothesis can be put forward that fuzzy algorithms for these images may work better than other algorithms. We designed an experiment based on the "plants in the greenhouse" image for two algorithms, K-Means and fuzzy C-means, and run 30 times. The box plot of RMSE values is presented in Figure 13.

Regarding the Figure 13, the hypothesis that the fuzzy algorithms work better for images with uniformly distributed colors is rejected.

*Table 4: The image histograms for three different images*

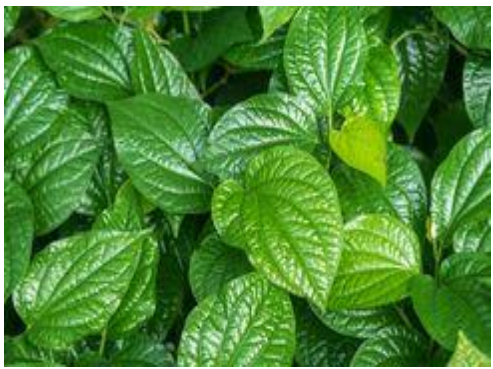

(a) Green leaves

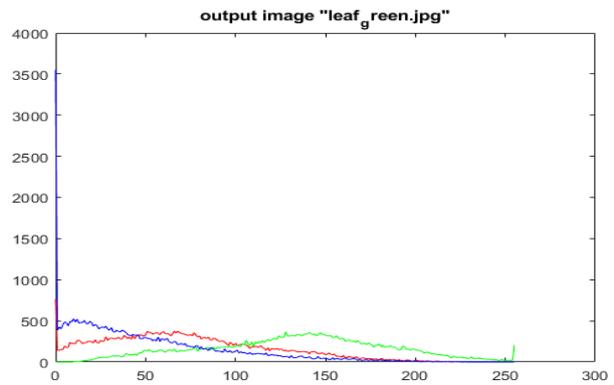

(b) Image histogram of Green leaves

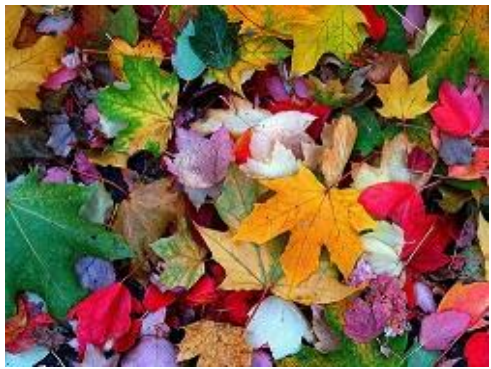

(c) Color leaves

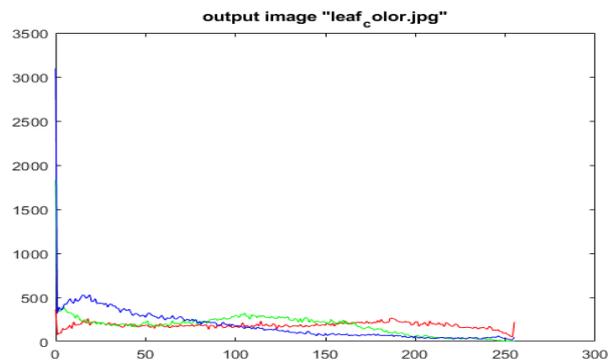

(d) Image histogram of Color leaves

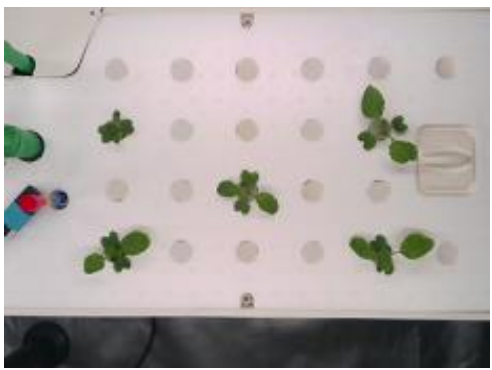

(e) Plants in the greenhouse

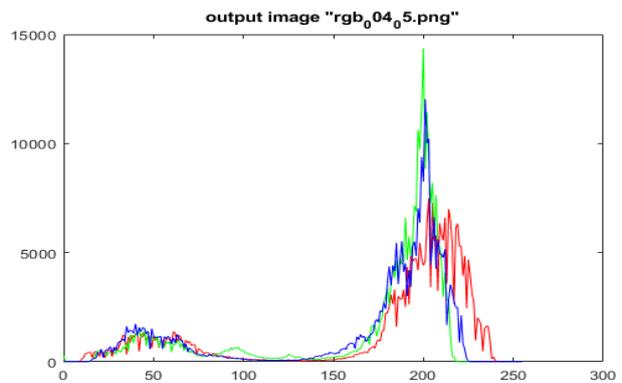

(f) Image histogram of Plants in the greenhouse

*Table 5: The RMSE, PSNR and SSIM from compressed images by K-Means++ in comparison with other algorithms*

|  | RMSE | | | | PSNR | | | | SSIM | | | |
|---|---|---|---|---|---|---|---|---|---|---|---|---|
|  | *p*-value | | Defeated by K-Means++ | | *p*-value | | Defeated by K-Means++ | | *p*-value | | Defeated by K-Means++ | |
|  | Gray | RGB | Gray | RGB | Gray | RGB | Gray | RGB | Gray | RGB | Gray | RGB |
| **K-means** | 3.5888e-04 | 6.9838e-06 | ✓ | ✓ | 3.0650e-04 | 7.6909e-06 | ✓ | ✓ | 5.3022e-05 | 4.0467e-05 | ✓ | ✓ |
| **fuzzy C-means** | 1.7344e-06 | 1.7344e-06 | ✓ | ✓ | 1.6341e-06 | 1.6331e-06 | ✓ | ✓ | 1.7333e-06 | 1.7131e-06 | ✓ | ✓ |
| **fuzzy C-means++** | 2.8308e-04 | 2.3704e-05 | ✓ | ✓ | 2.5217e-04 | 2.5967e-05 | ✓ | ✓ | 0.0010 | 2.0468e-04 | ✓ | ✓ |

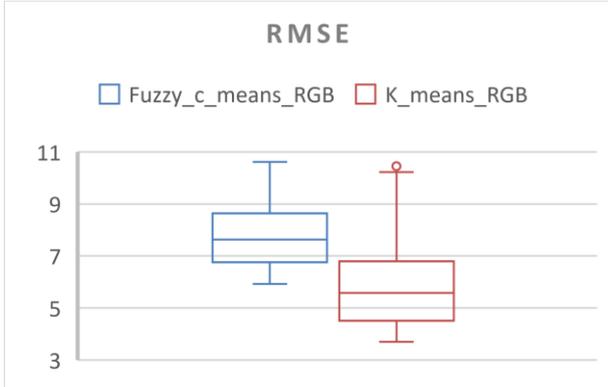

*Figure 13: The box plot of RMSE values for K-Means and fuzzy C-means algorithms*

### 4.7. Assessing the results by Statistical Tests

In this research, Wilcoxon signed-rank test [27] is used to assess the aforementioned algorithms. It is a non-parametric statistical hypothesis test applied to compare hypervolume of two equivalent samples from different algorithms in order to determine whether two dependent samples have similar distribution. The p-value of Wilcoxon signed rank test at the 5% significance level for RMSE, PSNR and SSIM metrics of compressed images by three algorithms (i.e. K-Means, fuzzy C-means, and fuzzy C-means++ ) based on K-Means ++ algorithm is prepared in Table 5. The p-value below the significance level shows that two desired algorithms are not from the same distribution and completely different. As indicated in Table 5, the p-value in all cases (i.e. grayscale and RGB) for all three metrics is less than the significance level, so, null hypothesis is rejected and shows that K-Means ++ results are significantly different from aforementioned algorithms results and overcomes them.

The reason why Fuzzy C-means algorithm showed poor results is that fuzzy logic is more applicable for rule based problems. However, image compression problem is not in that category.

## 5. Conclusion

IIoT is growing, day by day. One of the main applications of this technology is sending sensor images. These images typically have a high volume. So there is a need for a technique that reduces the volume without sacrificing quality. In this research, a new compression method to compact sensor-generated images in IIoT by using K-Means++ and Intelligent Embedded Coding (IEC), as our novel approach, was presented. This new technique was based on the colour pixels in a way that pixels with similar colours were clustered around a centroid, and in the next step, the colour of the pixels would be encoded. Compared with existing techniques that consistently compress and send images, the proposed method operated optimally and compressed and transmitted an image only if it differed from the prior sent image. The experiments were done based on an image dataset from a real smart greenhouse. The evaluation results showed that the new method could compress images with higher quality in comparison with other methods such as fuzzy C-means and K-means and fuzzy C-means++ clustering.

## 6. References

[1] J. M. Shapiro, "Embedded image coding using zerotrees of wavelet coefficients," *IEEE Trans. Signal Process.*, vol. 41, no. 12, pp. 3445–3462, Dec. 1993, doi: 10.1109/78.258085.

[2] N. Ahmed, T. Natarajan, and K. R. Rao, "Discrete Cosine Transform," *IEEE Trans. Comput.*, vol. C–23, no. 1, pp. 90–93, Jan. 1974, doi: 10.1109/T-C.1974.223784.

[3] Y. A. Ur Rehman, M. Tariq, and T. Sato, "A Novel Energy Efficient Object Detection and Image Transmission Approach for Wireless Multimedia Sensor Networks," *IEEE Sens. J.*, vol. 16, no. 15, pp. 5942–5949, Aug. 2016, doi: 10.1109/JSEN.2016.2574989.

[4] P. K. K and G. E, "Quality enhancement with fault tolerant embedding in video transmission over WMSNs in 802.11e WLAN," *Ad Hoc Netw.*, vol. 88, pp. 18–31, May 2019, doi: 10.1016/j.adhoc.2018.12.013.

[5] J. Paek and J. Ko, "$K$-Means Clustering-Based Data Compression Scheme for Wireless Imaging Sensor Networks," *IEEE Syst. J.*, vol. 11, no. 4, pp. 2652–2662, Dec. 2017, doi: 10.1109/JSYST.2015.2491359.

[6] Y. Song, H. Shin, and J. Paek, "Lightweight Server-Assisted H-K Compression for Image-Based Embedded Wireless Sensor Network," *IEEE Syst. J.*, vol. 13, no. 2, pp. 1386–1396, Jun. 2019, doi: 10.1109/JSYST.2018.2826004.

[7] J. Ahn, J. Park, D. Park, J. Paek, and J. Ko, "Convolutional neural network-based classification system design with compressed wireless sensor network images,"

*PLOS ONE*, vol. 13, no. 5, p. e0196251, May 2018, doi: 10.1371/journal.pone.0196251.

[8] S. Heng, C. So-In, and T. G. Nguyen, "Distributed Image Compression Architecture over Wireless Multimedia Sensor Networks," *Wirel. Commun. Mob. Comput.*, vol. 2017, 2017, doi: 10.1155/2017/5471721.

[9] R. Sheeja and J. Sutha, "Soft fuzzy computing to medical image compression in wireless sensor network-based tele medicine system," *Multimed. Tools Appl.*, Feb. 2019, doi: 10.1007/s11042-019-7223-2.

[10] M. A. Kaljahi, P. Shivakumara, M. Y. I. Idris, M. H. Anisi, and M. Blumenstein, "A new image size reduction model for an efficient visual sensor network," *J. Vis. Commun. Image Represent.*, vol. 63, p. 102573, Aug. 2019, doi: 10.1016/j.jvcir.2019.102573.

[11] J. Macqueen, "Some methods for classification and analysis of multivariate observations," in *In 5-th Berkeley Symposium on Mathematical Statistics and Probability*, 1967, pp. 281–297.

[12] J. C. Dunn, "A Fuzzy Relative of the ISODATA Process and Its Use in Detecting Compact Well-Separated Clusters," *J. Cybern.*, vol. 3, no. 3, pp. 32–57, Jan. 1973, doi: 10.1080/01969727308546046.

[13] J. C. Bezdek, *Pattern Recognition with Fuzzy Objective Function Algorithms*. Springer US, 1981. Accessed: Apr. 26, 2019. [Online]. Available: https://www.springer.com/gp/book/9781475704525

[14] D. Arthur and S. Vassilvitskii, "k-means++: The advantages of careful seeding," in *Proceedings of the eighteenth annual ACM-SIAM symposium on Discrete algorithms*, 2007, pp. 1027–1035.

[15] M. N. Reza, I. S. Na, S. W. Baek, and K.-H. Lee, "Rice yield estimation based on K-means clustering with graph-cut segmentation using low-altitude UAV images," *Biosyst. Eng.*, vol. 177, pp. 109–121, Jan. 2019, doi: 10.1016/j.biosystemseng.2018.09.014.

[16] J. Chen, M. Tian, X. Qi, W. Wang, and Y. Liu, "A solution to reconstruct cross-cut shredded text documents based on constrained seed K-means algorithm and ant colony algorithm," *Expert Syst. Appl.*, vol. 127, pp. 35–46, Aug. 2019, doi: 10.1016/j.eswa.2019.02.039.

[17] G. Sun, X. Liu, S. Wang, L. Gao, and M. Liu, "Width measurement for pathological vessels in retinal images using centerline correction and k-means clustering," *Measurement*, vol. 139, pp. 185–195, Jun. 2019, doi: 10.1016/j.measurement.2019.03.005.

[18] W. Rao, J. Xia, W. Lyu, and Z. Lu, "Interval data-based k-means clustering method for traffic state identification at urban intersections," *IET Intell. Transp. Syst.*, vol. 13, no. 7, pp. 1106–1115, Mar. 2019, doi: 10.1049/iet-its.2018.5379.

[19] M. Goyal and S. Kumar, "Improving the Initial Centroids of k-means Clustering Algorithm to Generalize its Applicability," *J. Inst. Eng. India Ser. B*, vol. 95, no. 4, pp. 345–350, Dec. 2014, doi: 10.1007/s40031-014-0106-z.

[20] K. A. A. Nazeer, S. D. M. Kumar, and M. P. Sebastian, "Enhancing the K-means Clustering Algorithm by Using a O(n logn) Heuristic Method for Finding Better Initial Centroids," in *2011 Second International Conference on Emerging Applications of Information Technology*, Feb. 2011, pp. 261–264. doi: 10.1109/EAIT.2011.57.

[21] M. S. Mahmud, M. M. Rahman, and M. N. Akhtar, "Improvement of K-means clustering algorithm with better initial centroids based on weighted average," in *2012 7th International Conference on Electrical and Computer Engineering*, Dec. 2012, pp. 647–650. doi: 10.1109/ICECE.2012.6471633.

[22] Fang Yuan, Zeng-Hui Meng, Hong-Xia Zhang, and Chun-Ru Dong, "A new algorithm to get the initial centroids," in *Proceedings of 2004 International Conference on Machine Learning and Cybernetics (IEEE Cat. No.04EX826)*, Aug. 2004, vol. 2, pp. 1191–1193 vol.2. doi: 10.1109/ICMLC.2004.1382371.

[23] K. Chowdhury, D. Chaudhuri, A. K. Pal, and A. Samal, "Seed selection algorithm through K-means on optimal number of clusters," *Multimed. Tools Appl.*, vol. 78, no. 13, pp. 18617–18651, Jul. 2019, doi: 10.1007/s11042-018-7100-4.

[24] S. Mittal and J. S. Vetter, "A Survey Of Architectural Approaches for Data Compression in Cache and Main Memory Systems," *IEEE Trans. Parallel Distrib. Syst.*, vol. 27, no. 5, pp. 1524–1536, May 2016, doi: 10.1109/TPDS.2015.2435788.

[25] H. Uchiyama *et al.*, "An easy-to-setup 3D phenotyping platform for KOMATSUNA dataset," in *Proceedings of the IEEE International Conference on Computer Vision*, 2017, pp. 2038–2045.

[26] W. Burger and M. J. Burge, *Digital Image Processing: An Algorithmic Introduction Using Java*. Springer, 2016.

[27] F. Wilcoxon, S. K. Katti, and R. A. Wilcox, "Critical values and probability levels for the Wilcoxon rank sum test and the Wilcoxon signed rank test," *Sel. Tables Math. Stat.*, vol. 1, pp. 171–259, 1970.